\documentclass[
a4paper,
aps,
prd,
preprint,
preprintnumbers,
nofootinbib,
]{revtex4-1}

\usepackage{fullpage}
\usepackage{amsfonts}
\usepackage{amsmath}
\usepackage{slashed}
\usepackage{amssymb}
\usepackage{graphicx}
\usepackage{epic}
\usepackage{eepic}
\usepackage{epsfig}
\usepackage{latexsym}
\usepackage{color}
\usepackage{float}
\usepackage{multirow}
\usepackage{hyperref}
\usepackage[caption=false]{subfig}
\usepackage{natbib}

\usepackage{shorthand}

\newcommand{\dmu}{\partial_\mu}
\newcommand{\dnu}{\partial_\nu}
\newcommand{\fmn}{F_{\mu\nu}}
\newcommand{\fmnup}{F^{\mu\nu}}
\newcommand{\hmn}{\widehat{F}_{\mu\nu}}
\newcommand{\hmnup}{\widehat{F}^{\mu\nu}}

\newcommand{\half}{\frac{1}{2}}

\newcommand{\ubr}[1]{\raisebox{1.5ex}{\hspace{#1ex}$\frown$\relax}}

\begin{document}

\preprint{UUITP-01/16}
~\\
\title{Heavy photophilic scalar at the LHC from a varying electromagnetic coupling}

\author{Ulf Danielsson}
\email{ulf.danielsson@physics.uu.se}

\author{Rikard Enberg}
\email{rikard.enberg@physics.uu.se}

\author{Gunnar Ingelman}
\email{gunnar.ingelman@physics.uu.se}

\author{Tanumoy Mandal}
\email{tanumoy.mandal@physics.uu.se}
\affiliation{Department of Physics and Astronomy, Uppsala University, Box 516, SE-751 20 Uppsala, Sweden}

\date{\today}

\begin{abstract}
We investigate the phenomenology of a heavy scalar $\phi$ of the type involved in Bekenstein's framework for varying electromagnetic coupling theories, with the difference that the 
scalar in our model has a large mass. The model has only two free parameters, the mass 
$M_{\phi}$ of the scalar and the scale $\Lambda$ of new physics.
The scalar is dominantly produced through photon-photon fusion at the LHC and leads to a diphoton final state. It can also be produced by quark-antiquark fusion in association with a 
photon or a fermion pair. Its dominating decay is to diphotons, but it also has a large 
three-body branching to a fermion pair and a photon, which can provide an interesting search 
channel with a dilepton-photon resonance. We derive exclusion limits on the 
$M_{\phi}-\Lambda$ plane from the latest 13 TeV LHC diphoton resonance search data. 
For a benchmark mass of $M_{\phi}\sim 1$ TeV, we find a lower limit on $\Lambda$ of 18 TeV. We discuss the more complex possibility of varying couplings in the full electroweak 
theory and comment on the possibility that the new physics is related to extra dimensions 
or string theory.
\end{abstract}



\maketitle

\section{Introduction}
\label{sec:intro}

In this paper we perform the first ever study of the collider physics of a model for space-time-varying gauge couplings. The model we study was constructed almost thirty-five years ago by Jacob Bekenstein~\cite{Bekenstein:1982eu,Bekenstein:2002wz} and introduces a new scalar field associated with a variation in the electromagnetic (EM) coupling constant ($\al_\text{EM}$). This model has, to the best of our knowledge, previously only been studied in the context of cosmology with a massless, or very light, scalar, and bounds on the model have been considered based on low-energy physics and astrophysics, see e.g.~\cite{Uzan:2010pm}. Here we propose that the Bekenstein model can be relevant for particle physics experiments, and that the scalar can have a mass on the TeV scale and therefore be accessible at the LHC.

Bekenstein's model is the first consistent such model -- it is Lorentz, gauge and time-reversal invariant and respects causality. The original motivation was to accommodate possible variations of the fine-structure constant over cosmological scales, but despite careful searches such variations have not been detected~\cite{Uzan:2010pm}. 

In string theory all couplings are associated with scalar fields called moduli, and are thus subject to variations. The excitations of these fields are typically very heavy and as a consequence the couplings will be locked to essentially constant values, given by the specific compactification scenario, see e.g.~\cite{Zagermann:2011asa} for a discussion. The Bekenstein model is not derived from string theory, but we consider it the simplest consistent scenario for associating couplings with scalar fields. 

Considering variations of $\alpha_\text{EM}$, this implies the existence of a new scalar field $\phi$ which couples to photons, i.e., it is  “photophilic”. This would provide a discovery potential at the LHC through the decay $\phi\to \gamma \gamma$, which gives a striking signal of high-energy photons pairs with invariant mass $M_{\gamma\gamma}=M_{\phi}$ that may extend to several TeV. The possibility of such a discovery signal was illustrated by the excesses in the diphoton invariant mass distribution around 750 GeV reported~\cite{ATLAS:2015dxe,CMS:2015dxe} by both ATLAS and CMS based on the first 13 TeV data with $\sim 3$ fb$^{-1}$ integrated luminosity, even though no excess was observed when the larger dataset of $\sim 16$ fb$^{-1}$ had later been collected~\cite{ATLAS:2016eeo,Khachatryan:2016yec}. 

Although no signal for such a scalar is presently observed, it is well worth investigating
this theory for Beyond Standard Model (BSM) physics which is fundamentally different from other BSM theories considered at the LHC. Therefore, we will here derive an explicit model for such a new scalar field $\phi$ and study the general phenomenology at the LHC. In particular, we use the available data to constrain the model parameters. The model is economic in the sense that it introduces only one new field and two new parameters, i.e., the scalar $\phi$ with mass $M_{\phi}$, and the new energy scale $\Lambda$. With $M_{\phi}$ around the electroweak (EW) or TeV scale, $\alpha_\text{EM}$ remains constant throughout most of the history of the universe and no deviations would have been detectable so far.\footnote{The new scalar should also be investigated as a candidate for the inflaton. The quadratic contribution to the potential, with a TeV scale mass cannot be used for inflation, but one might consider higher order terms providing a useful shape.} 

The “photophilic” nature of this new scalar favors its production through 
photon-photon ($\gm\gm$) fusion rather than through gluon-gluon ($gg$) fusion as is more commonly considered for new scalars in BSM theories. Although the scalar in our model can be produced in a pure $s$-channel process, the basic new vertices also give production modes together with a photon or a dijet or dilepton pair, which imply characteristic predictions of the model to be tested against data.
 
This paper is organized as follows. In Section \ref{sec:model} we introduce the model and derive the interactions of the new scalar. We also discuss briefly the generalization to the whole EW sector of the SM. In Section \ref{sec:PDLHC} we discuss the phenomenology of the model, and through detailed numerical comparison with LHC limits we extract bounds on the model parameters in Section \ref{sec:exclu}. In Section \ref{sec:conclusions} we conclude by considering the results in a larger context.

\section{The model}
\label{sec:model}

In the Bekenstein model \cite{Bekenstein:1982eu,Bekenstein:2002wz} the variation of the coupling $e$ is derived from an action that reduces to electromagnetism for constant coupling. It is assumed that the space-time variation of the  coupling is given by $e = e_0 \epsilon(x)$, where $\epsilon(x)$ is a scalar field with dynamics given by the kinetic term
\begin{equation}
\frac{1}{2}\frac{\Lambda^2}{\epsilon^2} (\dmu\epsilon)^2,
\end{equation}
where $\Lambda$ is an energy scale. It is assumed that the field $\epsilon$ multiplies the electric charge $e$ everywhere in the Lagrangian of the model. Specifically, this means that $eA_\mu$ is everywhere replaced by $e_0 \epsilon A_\mu$. Gauge invariance, and invariance with respect to a rescaling of $\epsilon$, then requires that the field strength tensor be given by
\begin{equation}
\fmn = \frac{1}{\epsilon} \left[ \dmu (\epsilon A_\nu) - \dnu (\epsilon A_\mu) \right].
\end{equation}
To keep our notation as explicit as possible we define $\hmn=\dmu A_\nu - \dnu A_\mu$, and introduce a scalar field $\varphi$ such that $\epsilon = e^\varphi$. We will be assuming small fields, working at lowest order, and therefore write  $\epsilon \simeq 1+\varphi$, and keep only terms linear in $\varphi$. We then find, for the kinetic term of the  field, that $\fmn\fmnup = \hmn\hmnup + 4 \, \dmu\varphi \, A_\nu  \hmnup $. Finally, we define a new field $\phi=\varphi \Lambda$ so that all fields have their usual mass dimensions. In this way we find a  Lagrangian for electromagnetism plus the scalar field given by
\begin{equation}
 {\cal L} \supset \half(\dmu\phi)^2
-\frac{1}{4} \hmn\hmnup  - \frac{1}{\Lambda} \dmu\phi \, A_\nu  \hmnup .
\label{eq:scalarlagr}
\end{equation}

\begin{figure}[h!]
\includegraphics[scale=1]{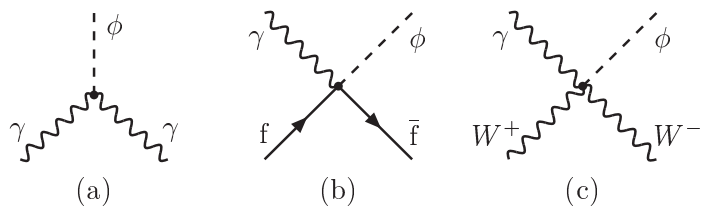}
\caption{Basic interaction vertices of the scalar field $\phi$ as given by the Lagrangian in Eq.~(\ref{eq:Lag}).}
\label{fig:vertices}
\end{figure}

Because of the definition of $e$, the new scalar field $\phi$ will in addition to the interaction in Eq.~(\ref{eq:scalarlagr}) couple to all electrically charged fields. The coupling of the EM field to charged fermions is obtained from the covariant derivative, which is given by $D_\mu = \dmu - i e Q A_\mu$, where $Q$ is the charge of the coupled field, and $e=e_0\epsilon$ so that $D_\mu = \dmu - i e_0 Q A_\mu - i e_0 Q (\phi/\Lambda) A_\mu$.
If we now define $\widehat D_\mu = \dmu- i e_0 Q A_\mu$ as the more familiar covariant derivative of electromagnetism,
the gauge invariant kinetic term for fermions is given by
\begin{equation}
{\cal L} \supset i\overline\psi \slashed{D} \psi
= i\overline\psi \widehat{\slashed{D}} \psi + \frac{e_0 Q}{\Lambda} \phi\, \overline\psi \gamma^\mu \psi \, A_\mu .
\end{equation}
The final interaction of $\phi$ to consider is the coupling to $W^\pm$ bosons, which is obtained by inserting $eA_\mu=e_0\epsilon A_\mu$ in the EW gauge kinetic term written in terms of the mass eigenstates. The resulting terms, to lowest order in the electric coupling, are shown below in Eq.~(\ref{eq:Lag}). There are also five-point vertices of the type $\phi VV W^+W^-$ coming from the quartic gauge boson couplings. These couplings contain a factor $e_0^2/\Lambda^2$ and are suppressed with respect to the displayed couplings. For instance, there are four-body decays of $\phi$, such as $\phi\to W^+W^-\gamma\gamma$ where the five-body vertices contribute. The amplitude from the vertices in Fig.~\ref{fig:vertices} is proportional to $e_0^2/\Lambda$, while the amplitude from the $\phi \gamma\gamma W^+W^-$ vertex is proportional to $e_0^2/\Lambda^2$. For this reason we neglect the higher-order terms in this paper.

It is shown in~\cite{Bekenstein:1982eu}, see also \cite{Magueijo:2001ir}, that one through a partial use of the equation of motion for $F_{\mu \nu}$ can map the theory to a new theory where the scalar only couples through $(1/2\Lambda) \phi \hmn\hmnup $, without any coupling between the scalar and the fermions.  These theories are equivalent, and we here consider the original action due to Bekenstein.

As we have described above, this model has been used as a framework for a space-time varying 
$\alpha_\text{EM}$ with a massless or very light scalar field $\phi$. We will now add one new ingredient: a mass term for the scalar field with $M_\phi$ around the EW scale or up to several TeV. The energy cost of moving $\phi$ away from its minimum will then be very large for energies $\lesssim M_\phi$, so that $\alpha_\text{EM}$ will have negligible variation.

To get the Lagrangian in its final form, let us now drop the hats on the above expressions, so that from now on,
$\fmn$ is the standard field strength tensor. Collecting all terms, we then have the Lagrangian of our model
\begin{align}
{\cal L} = {\cal L}_\text{SM} &+ \half(\dmu\phi)^2 - \half M_{\phi}^2\phi^2   - \frac{1}{\Lambda} \dmu\phi \, A_\nu  \fmnup
 + \frac{e_0 Q}{\Lambda} \phi\, \overline\psi \gamma^\mu \psi \, A_\mu
\nonumber \\
&+
\frac{ie_0}{\Lambda} \phi \left[
W_{\mu\nu}^{+}{W^{-}}^{\mu}A^{\nu} -
W_{\mu\nu}^{-}{W^{+}}^{\mu}A^{\nu} + F^{\mu\nu}  W^{+}_{\mu}W^{-}_{\nu}  \right]
\nonumber\\
&+
\frac{ie_0}{\Lambda}\left[(A_\nu \dmu\phi - A_\mu \dnu\phi)  W^{+}_{\mu}W^{-}_{\nu}
\right] + \mathcal{O}\lt(\Lm^{-2}\rt), \label{eq:Lag}
\end{align}
where ${\cal L}_\text{SM}$ represents the ordinary Standard Model Lagrangian, 
$W_{\mu\nu}^{\pm}=\dmu W^{\pm}_{\nu}-\dnu W^{\pm}_{\mu}$ and $\psi$ is a generic field with charge $Q$ denoting all electrically charged fermions of the Standard Model, written as Dirac spinors for both left- and right-handed components. 
Thus our model has only two new parameters, the mass $M_\phi$ and the scale $\Lambda$. Note that the EM coupling $e_0$ in Eq.~(\ref{eq:Lag}) is the usual not varying coupling, and the dynamics of the varying constant now sit instead in the scalar field $\phi$.\footnote{For example, at very high energies above the mass of $\phi$, small variations of the coupling constant $e$ could become visible in $e^+e^-$ collisions at a future linear collider. In the formulation here, where we have the fixed constant $e_0$, such variations would be associated with loops or real or virtual emission diagrams involving $\phi$.}

The vertices are shown in Fig.~\ref{fig:vertices} and the corresponding Feynman rules are
\begin{eqnarray}
\text{Fig.~\ref{fig:vertices}a} &: \quad \gamma\gamma\phi &\to \,
\frac{i}{\Lambda} \left(p_{1\mu} p_{3\nu} + p_{3\mu} p_{2\nu} + g_{\mu\nu} \, p_3\cdot p_3 \right)
\label{ggphi_rule}\\
\text{Fig.~\ref{fig:vertices}b} &: \quad {\rm f} \bar{\rm f}\gamma\phi &\to \,
\frac{ie_0Q}{\Lambda} \gamma^\mu
\label{ffAphi_rule}\\
\text{Fig.~\ref{fig:vertices}c} &: \quad \gamma\phi W^+ W^- &\to \,
\frac{i e_0}{ \Lambda } \left(
  p_1^{\mu_3} g_{\mu_1,\mu_4}
- p_1^{\mu_4} g_{\mu_1,\mu_3}
+ p_2^{\mu_3} g_{\mu_1,\mu_4}
- p_2^{\mu_4} g_{\mu_1,\mu_3} \right. \nonumber\\
&&\quad +  \left.
  p_3^{\mu_4} g_{\mu_1,\mu_3}
- p_3^{\mu_1} g_{\mu_3,\mu_4}
+ p_4^{\mu_1} g_{\mu_3,\mu_4}
- p_4^{\mu_3} g_{\mu_1,\mu_4}
\right) ,
\label{WWAphi_rule}
\end{eqnarray}
where in (\ref{ggphi_rule}) and (\ref{WWAphi_rule}) the numbered indices refer to the particles in the order listed.

Above, we have only let the EM coupling $e$ vary. This has the consequence that only the photon and the fields carrying electric charge couple to 
$\phi$. It is possible to instead let the $\mathrm{U(1)}_Y$ or 
$\mathrm{SU(2)}_L$ or both couplings vary, which will couple the 
$\phi$ not only to the photons but to the weak gauge bosons as well.
The Bekenstein model of varying-$\alpha_\text{EM}$ theory has been generalized~\cite{Kimberly:2003rz} to vary the 
$\mathrm{SU(2)}_L$ and $\mathrm{U(1)}_Y$ couplings of the electroweak (EW) theory, either with both couplings varying in the same way (one scalar field associated with
both variations), or independently of each other. In the latter case, there will be two scalar fields associated with the variations of the two gauge couplings.
 
Let us therefore here briefly consider the extension of the model to the entire electroweak gauge group. 
Consider, therefore, two scalars $S$ and $S'$ associated with the variation of 
$g$ and $g'$, where $g$ and $g'$ are the $\mathrm{SU(2)}_L$ and $\mathrm{U(1)}_Y$ gauge couplings respectively. The interaction Lagrangian (before EWSB) can be expressed as (for simplicity using the equivalent formulation discussed in connection with Eq.~\eqref{eq:scalarlagr})
\be
\label{eq:intlag}
\mc{L} \supset  \frac{1}{2\Lm}S B_{\mu\nu}B^{\mu\nu} + \frac{1}{2\Lm'}S' W_{\mu\nu}W^{\mu\nu}\ ,
\ee
where $B_{\mu\nu}$ and $W_{\mu\nu}$ are the field-strength tensors and 
$\Lm$ and $\Lm'$ are the scales for the 
$\mathrm{SU(2)}_L$ and $\mathrm{U(1)}_Y$ gauge groups respectively. 
In Eq.~\eqref{eq:intlag}, we now replace $B_{\mu} = c_wA_{\mu}-s_wZ_{\mu}$ 
and $W^3_{\mu} = s_wA_{\mu}+c_wZ_{\mu}$ where $s_w$ and $c_w$ are the sine and 
cosine of the Weinberg angle, respectively. After this replacement, we obtain
\be
\label{eq:intlag2}
\mc{L} \supset \frac{1}{2}\lt(c_w^2\frac{S}{\Lm} + s_w^2\frac{S'}{\Lm'}\rt)F_{\mu\nu}F^{\mu\nu} + \frac{1}{2}\lt(s_w^2\frac{S}{\Lm} + c_w^2\frac{S'}{\Lm'}\rt)Z_{\mu\nu}Z^{\mu\nu} - s_wc_w\lt(\frac{S}{\Lm}-\frac{S'}{\Lm'}\rt)F_{\mu\nu}Z^{\mu\nu} .
\ee
The fields $S$ and $S'$ need not be the mass eigenstates, which in general could be linear combinations of $S$ and $S'$.
We therefore define the mass eigenstates $\phi$ and $\phi'$ as the linear combinations
of $S$ and $S'$ 
\be
\label{eq:lincomb} 
\phi = \cos\al~S + \sin\al~S';~~~ \phi' = -\sin\al~S + \cos\al~S' \ , 
\ee
where $\al$ is the mixing angle  determined from the free
parameters of the scalar potential (which we will not present here). We may
identify $\phi$ as the lightest scalar and $\phi'$ as a heavier mass eigenstate 
by properly arranging various free parameters in the scalar potential.
In the following, we only focus on the lightest scalar, which may reasonably be expected
to first show up at the LHC, and assume that signatures of the
heavier will be detected later.\footnote{Of course, the two scalars may also be close in mass.
We will consider this possibility in the future.} In this general set-up, both $\phi$ and 
$\phi'$ will decay to $\gm\gm$, $\gm Z$, $ZZ$ and $WW$ modes. In case $\Lm'\gg \Lm\sim \mc{O}(1)$ TeV, the BR $\phi\to WW$ becomes suppressed compared to other BRs of $\phi$. Neglecting the small 
phase-space suppression for the heavy gauge bosons in the final state, the two-body
branching ratios (BRs) of $\phi$ are in the following proportions,
\be 
\mathrm{BR}_{\gm\gm}:\mathrm{BR}_{\gm Z}:\mathrm{BR}_{ZZ} \approx
\frac{1}{2}\lt(\frac{c_w^2\cos\al}{\Lm} \rt)^2:s_w^2c_w^2 \lt(\frac{\cos\al}{\Lm} \rt)^2:
\frac{1}{2}\lt(\frac{s_w^2\cos\al}{\Lm} \rt)^2\ .
\ee
Therefore, for any values of $\al$ and $\Lm$, the BRs of $\phi$ becomes (assuming $\Lm'\gg \Lm$)
\be
\label{eq:BREW}
\mathrm{BR}_{\gm\gm}:\mathrm{BR}_{\gm Z}:\mathrm{BR}_{ZZ} \approx \frac{1}{2}c_w^4:
s_w^2c_w^2:\frac{1}{2}s_w^4 \approx 60\%:35\%:5\%
\ee
One can see that the $\gm\gm$ decay of $\phi$ dominates and the $\gm Z$
and $ZZ$ decays are suppressed and therefore not observable with presently available data samples.
In this first study, it is
therefore well motivated to find the most essential phenomenology of this varying coupling
theory by the simplification to only vary the EM coupling and concentrate on final states with photons that provide clean experimental signals.
Thus, we leave the more complex study of the full theoretical framework for future work~\cite{inprep}.

\section{Decays and Production at the LHC}
\label{sec:PDLHC}

In this section, we study the phenomenology of our model and derive limits on the
two model parameters $M_{\phi}$ and $\Lm$ from the relevant 8 and 13 TeV data from the LHC.
In particular, we use $\gm\gm$~\cite{ATLAS:2016eeo,Khachatryan:2016yec}, 
$\gm Z$~\cite{CMS:2016pax,CMS:2016cbb} and $ZZ$~\cite{ATLAS:2016npe} resonance searches
from the LHC. The $\gm\gm$ data are used for the simplified model where $\phi$ only couples 
to photons, but all three types of data are used to set limits on the varying EW theory. 
We implement the Lagrangian shown in Eq.~(\ref{eq:Lag}) in {\sc FeynRules2}~\cite{Alloul:2013bka} 
to generate the model files 
for the {\sc MadGraph5}~\cite{Alwall:2014hca} event generator. We use the 
{\sc MMHT14LO}~\cite{Harland-Lang:2014zoa} parton distribution functions (PDFs) to compute cross sections.
Generated events are passed through {\sc Pythia8}~\cite{Sjostrand:2014zea} for parton shower and hadronization. 
Detector simulation is performed for ATLAS and CMS using {\sc Delphes3}~\cite{deFavereau:2013fsa}
which uses the {\sc FastJet}~\cite{Cacciari:2011ma} package for jet clustering using the 
anti-$k_T$ algorithm~\cite{Cacciari:2008gp} with clustering parameter $R=0.4$. 

\begin{figure}[!t]
\subfloat[]{\includegraphics[scale=0.6]{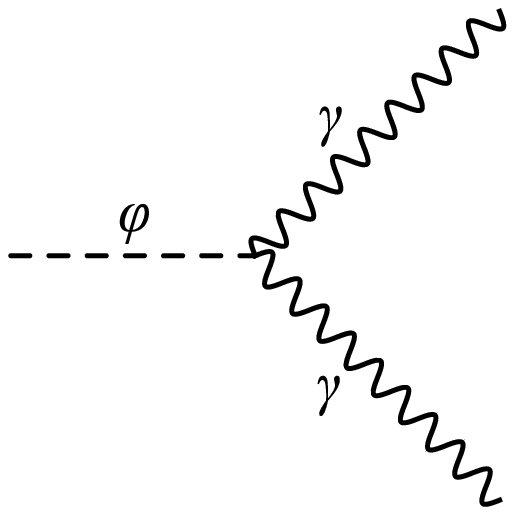}\label{fig:hp2aa}}\hspace{0.5cm}
\subfloat[]{\includegraphics[scale=0.6]{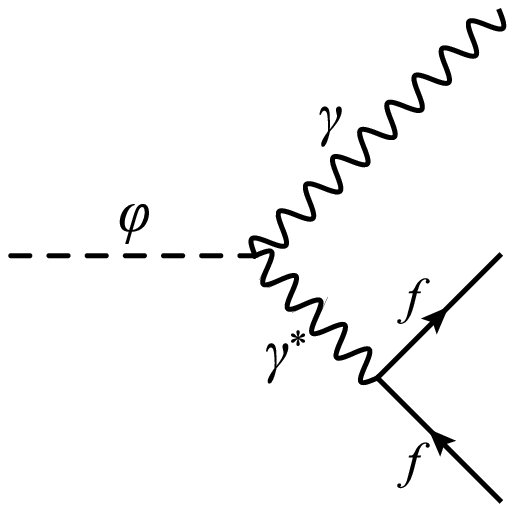}\label{fig:hp2aff2}}\hspace{0.5cm}
\subfloat[]{\includegraphics[scale=0.6]{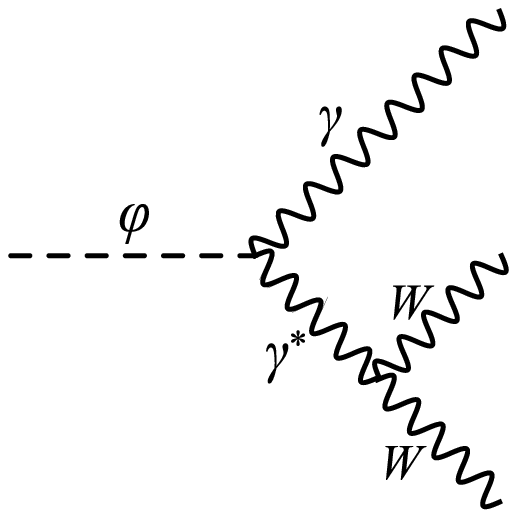}\label{fig:hp2ww2}}
\caption{Sample Feynman diagrams of the two and three-body decay modes of $\phi$.}
\label{fig:FDhpDecay}
\end{figure}

Since $\phi$ originates from the variation of the fine-structure constant, it directly 
couples to photons through an effective dimension-5 operator of the type $\phi F_{\mu\nu}^2$.
The only possible two-body decay of $\phi$ is the diphoton mode (Fig.~\ref{fig:hp2aa}), which is a tree level decay -- not a loop-induced
decay through a charged particle as for the SM Higgs. There are subdominant
three-body decays of $\phi$ mediated through an off-shell photon
as shown in Figs.~\ref{fig:hp2aff2} and \ref{fig:hp2ww2}. 
%
The partial decay width into two photons, shown in Fig.~\ref{fig:hp2aa}, is given by
\begin{equation}
\Gamma(\phi \to \gamma\gamma) = \frac{M_{\phi}^3}{16\pi \Lambda^2} .
\end{equation}
The analytical expressions for the subdominant three-body decay modes are more complicated 
due to massive particles in the three-body phase space, and we compute 
partial widths of those modes numerically in {\sc MadGraph5}.
In Table~\ref{tab:PWBR}, we show the partial widths and branching
ratios (BR) of $\phi$ into its two and three body decay modes for $M_{\phi}=1$ TeV and 
$\Lm=2$ TeV.
It is important to note that the BRs of $\phi$ only depend on its mass but are independent of the
scale $\Lm$. This is because all partial widths and hence the total width scale as $\Lm^{-2}$ and the $\Lm$
dependence cancels in the ratios.
From Table~\ref{tab:PWBR}, we can see that $\phi\to\gm\gm$ is the dominant decay mode and this mode has a branching ratio
of about 65\%, so BRs of $\phi$ to other modes are non-negligible. In our analysis we always use the total decay width including 
all contributions coming from the three body decays.


\begin{table}[!t]
\begin{tabular}{|c|c|c|c|c|}
\hline
Decay Mode & $\phi\to \gm\gm$ & $\phi\to \gm ff(jj)$ & $\phi\to \gm WW$ & Total \\
\hline
Width (GeV) & 5.0 & 1.9 (0.86) & 0.79 & 7.6 \\
\hline
BR (\%) & 65 & 25 (11) & 10 & - \\
\hline
\end{tabular}
\caption{The partial widths and BRs of $\phi$ for $M_{\phi}=1$ TeV. The widths are proportional to $\Lm^{-2}$ and are here given for $\Lm=2$ TeV, whereas the BRs are independent
of $\Lm$. Here, $f$ includes all SM charged fermions and $j$
denotes jets of ``light" quarks, including $b$ quarks.}
\label{tab:PWBR}
\end{table}

\begin{figure}[b]
\subfloat[]{\includegraphics[height=3cm,width=3cm]{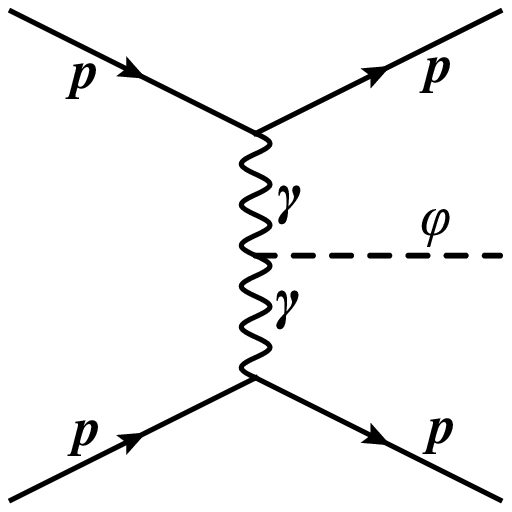}\label{fig:fullela}}\hspace{0.25cm}
\subfloat[]{\includegraphics[height=3cm,width=3cm]{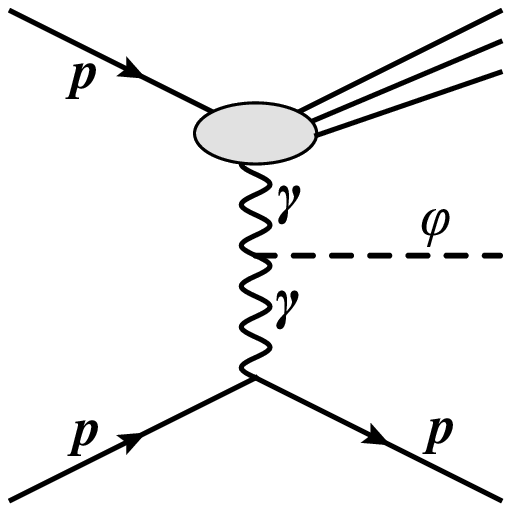}\label{fig:semiinela}}\hspace{0.25cm}
\subfloat[]{\includegraphics[height=3cm,width=3cm]{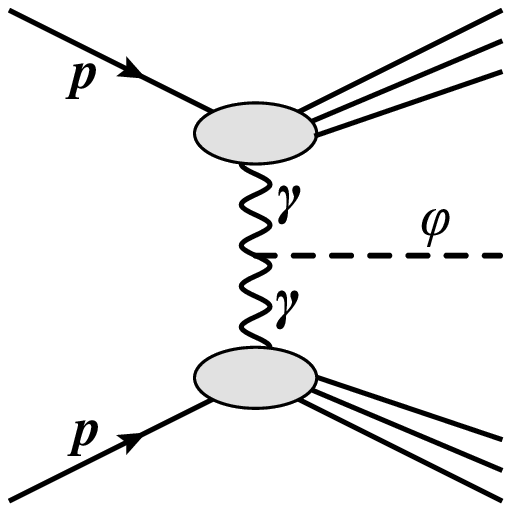}\label{fig:fullinela}}\hspace{0.25cm}
\subfloat[]{\includegraphics[height=3cm,width=3cm]{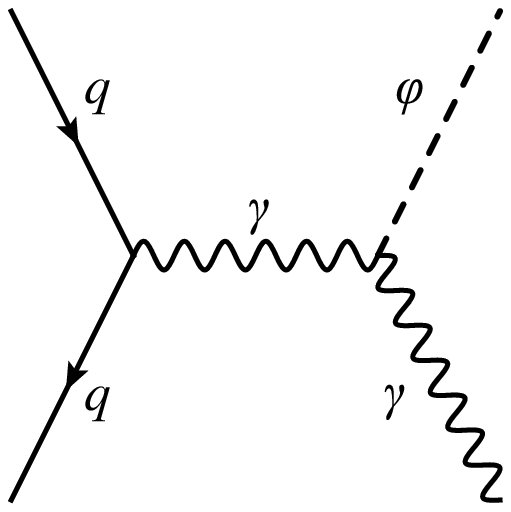}\label{fig:qq2hpa2}}
\caption{Sample Feynman diagrams of the production of $\phi$ at the LHC. (a), (b) and (c) are
the fully elastic, semi elastic and fully inelastic contributions to the $\gm\gm$ fusion
production respectively. (d) The quark-antiquark initiated production of $\phi$ in association with a photon.}
\label{fig:FDpp2hpx}
\end{figure}

In Fig.~\ref{fig:FDpp2hpx}, we show a few sample Feynman diagrams of the main production
channels of $\phi$ at the LHC.
Unlike the $gg$ initiated SM-like Higgs boson production, these channels are
induced by $\gm\gm$, $\gm q$ and $qq$ initial states. The dominant production channel
of $\phi$ is the $\gm\gm$ fusion where the initial photons come from the 
photon distribution of the proton. In the $\gm\gm$ fusion process, the
dominant contribution comes from the inelastic scattering, where the proton would 
break up. On the other hand, elastic collisions, where the protons remain intact,
are subdominant but provide a much cleaner channel that can be identified with forward detectors.
This exclusive production of $\phi$, i.e. $pp\to\phi pp$,
where two forward protons are tagged, is an interesting process to search for (this production mode has been considered in \cite{Fichet:2015vvy,Csaki:2015vek,Anchordoqui:2015jxc}, for models very different from ours).

In Table~\ref{tab:CS}, we present the partonic cross sections of
various production modes of $\phi$ for the 8 and 13 TeV LHC for $M_{\phi}=1$ TeV and $\Lm=2$ TeV.
To compute these cross sections we apply
the following basic kinematical cuts at the generation level wherever they are applicable:
\be
\label{eq:cut}
p_T(x)>25~{\rm GeV};~~|\eta(x)|<2.5;~~\Dl R(x,y)>0.4~~~{\rm where}~~x,y\equiv \{\gm,j\}
\ee

\begin{table}[!t]
\begin{tabular}{|c|c|c|c|c|c|c|}
\hline
Production mode & $\gm\gm\to\phi$ & $\gm p\to\phi j$ & $pp\to\phi jj$ & $pp\to\phi \gm$ & $pp\to\phi \gm j$ & $pp\to\phi \gm jj$  \\
\hline
CS@8TeV (fb)  & 32 & 7.8 & 0.45 & 0.18 & 0.10 & 0.04  \\
\hline
CS@13TeV (fb) & 110 & 30 & 1.8 & 1.1 & 0.71 & 0.40  \\
\hline
\end{tabular}
\caption{Partonic cross sections of various production channels of $\phi$ for $M_{\phi}=1$ TeV and $\Lm=2$ TeV computed at renormalization ($\mu_R$)
and factorization ($\mu_F$) scales $\mu_R=\mu_F=M_{\phi}=1$ TeV for the LHC at 8 and 13 TeV. These cross
sections are computed using MMHT14LO PDFs by applying some basic generation level cuts as defined in Eq.~(\ref{eq:cut}).
Here, $p$ includes $b$-quark PDF and $j$ denotes light jets including $b$-jet. All signal cross sections scale as $\Lambda^{-2}$.}
\label{tab:CS}
\end{table}

In our analysis, we include elastic, semi-elastic and inelastic contributions (as shown in 
Fig.~\ref{fig:FDpp2hpx}) in the
$\gm\gm$ fusion process. In order to include these three contributions properly without
double counting, we use the MLM matching algorithm~\cite{Mangano:2006rw} to match matrix element partons with the parton shower
to generate inclusive $pp\to\phi$ signal events.
Our inclusive signal includes up to two jets and we generate
it by combining the following processes,
\be
\left. \begin{array}{lclcl}
\gm\gm &\to &(\phi) 		&\to & \gm\gm\ubr{-2.5}\,,\\
\gm p &\to &(\phi\ j)		&\to & \gm\gm\ubr{-2.5}\ j\,,\label{eq:match}\\
pp &\to &(\phi\ jj)  	&\to & \gm\gm\ubr{-2.5}\ jj\,.
\end{array}\right\}
\ee
where we set the matching scale $Q_{cut}=125$ GeV and the curved connection above two photons
signify that they come from the decay of $\phi$. To determine the appropriate $Q_{cut}$
for the process, we check the smooth transition in differential jet rate distributions between events with 
$N$ and $N+1$ jets. Moreover, variations of $Q_{cut}$ around 125 GeV would not change the matched cross section much which we ensured to be within $\sim 10\%$ of the zero jet contribution. The matched cross section of the inclusive
(up to 2-jets) $pp\to\phi+{\rm jets}$ process is roughly 64 (20) fb for 13 (8) TeV LHC and
includes the $\phi\to\gm\gm$ BR. For a TeV-scale resonance the $\gm\gm\to\phi\to\gm\gm$ with
parton shower contribution is very similar to the total matched cross sections. Therefore,
one can use just $\gm\gm$ process in a simplified analysis for a high mass resonance.
We also generate inclusive $pp\to\phi\gm$ (up to 2-jets) events by using similar matching
technique. This channel has an interesting final state with three hard photons.
In most events, the first and second hardest photons come from $\phi$-decay, and their 
transverse momentum ($p_T$) distributions roughly peak around $M_{\phi}/2$, as they come from the 
decay of $\phi$. We observe that the third photon, although relatively softer than the first two, is also
moderately hard. When one constructs the invariant mass ($M$) of the two hardest photons, the third photon
would not contaminate much. As a result, we observe a sharp peak around $M_{\phi}$ in the 
invariant mass distribution of the first and second hardest photons $M(\gm\gm)$. Therefore, this channel
can provide an interesting signature with three hard photons with $M(\gm\gm)$ of first two
peaking at $M_{\phi}$.

There is another channel $pp\to\phi\to \gm\ell\ell$ (where $\ell=\{e,\mu\}$) which
might also be interesting to search for as the BR of $\phi$ to $\gm\ell\ell$ mode is relatively
big and also very clean final state. But since the two leptons are coming from an off-shell
photon, they are not spatially very separated and therefore a dedicated analysis is required
to observe this channel. Similar analyses for the SM Higgs have been done previously for the
LHC~\cite{Pozdnyakov:2016gbz}.

\section{Exclusion from the LHC data}
\label{sec:exclu}

Run-I and run-II LHC data on the diphoton resonance searches set strong upper limits
on $\sg\times{\rm BR}$. It should be remembered that 
all these searches are generally optimized for an $s$-channel resonance (spin-0 or spin-2)
produced through $gg$ fusion and decaying to two photons. 
For exclusive searches, one demands exactly two selected photons with no selected jet, whereas for inclusive 
searches, one keeps events with at least two selected photons and any number of selected jets.  
In order to derive a limit on $\Lm$ by recasting the $\sg\times{\rm BR}$ upper limit from an experiment, we need to properly take care
of the cut efficiencies. The cut efficiencies can significantly change
for different selection criteria and also for different production mechanisms.
For instance, in our case, the scalar is dominantly produced through $\gm\gm$ fusion
and the signal cut efficiency can be different from $gg$ fusion production.
This can be properly taken care of by using the following relation:
\be
\label{eq:Ns}
\mc{N}_s =\sg_s\times \ep_s\times \mc{L}=\lt(\sum_i \sg_i\times\ep_i\rt)\times\mc{L}\ ,
\ee
where $\mc{N}_s$ is the number of events for the signal considered for luminosity $\mc{L}$ and $\sg_s$ is the
corresponding signal cross section. The corresponding signal
cut efficiency is denoted by $\ep_s$. When different types of signal topology and/or
final state contribute to any experimental observable, $\mc{N}_s$ can be expressed by the sum
$\lt(\sum_i \sg_i\times\ep_i\rt)\times\mc{L}$. Here, $i$ runs over all contributing
signal processes to any observable.

We roughly employ the selection cuts used by ATLAS~\cite{ATLAS:2016eeo} and 
CMS~\cite{Khachatryan:2016yec}
for their 13 TeV analyses as listed below. Although they have not mentioned any jet selection,
we include basic jet selection cuts in the following list to demonstrate how our signal
cut efficiencies change with different number of selected photons and jets.
\bi

\item \underline{Selection cuts for ATLAS 13 TeV analysis:}

\ben
\item Transverse momentum of the selected photons and jets satisfy $p_T(\gm),p_T(j) > 25$ GeV.
\item Pseudorapidity of the selected photons satisfy $|\eta(\gm)|<2.37$ excluding
the barrel-endcap region $1.37 < |\eta(\gm)| < 1.52$ and jets $|\eta(j)|<4.4$.
\item Separation in $\eta$-$\phi$ plane between any two photons or photon-jet pair
satisfies $\Dl R(\gm,\gm),\Dl R(\gm,j) > 0.4$.
\item Invariant mass of the two hardest photons and their transverse momenta satisfy the
relations $p_T(\gm_1)/M(\gm_1,\gm_2)>0.4$ and $p_T(\gm_2)/M(\gm_1,\gm_2)>0.3$.
\een

\item \underline{Selection cuts for CMS 13 TeV analysis:}

\ben
\item Transverse momentum of the selected photons and jets satisfy $p_T(\gm),p_T(j) > 25$ GeV
with two hardest photons satisfying $p_T(\gm_1),p_T(\gm_2) > 75$ GeV.
\item Pseudorapidity of the selected photons satisfy $|\eta(\gm)|<2.5$ excluding
the barrel-endcap region $1.44 < |\eta(\gm)| < 1.57$ (and rejecting events with both photons
are in endcap region) and jets $|\eta(j)|<4.5$.

\item Separation in $\eta$-$\phi$ plane between any two photons or photon-jet pair
satisfies $\Dl R(\gm,\gm),\Dl R(\gm,j) > 0.4$.
\item Invariant mass of the two hardest photons satisfies $M(\gm_1,\gm_2)>230$ GeV and 
$M(\gm_1,\gm_2)>320$ GeV for events with at least one photon is in endcap region.
\een
\ei

Now, we want to investigate how selection cut efficiencies depend on the different selection 
criteria imposed on the number of photons and jets.
In Table~\ref{tab:CEhpa}, we show cut efficiencies for different selection
criteria on the number of photons and jets for the inclusive $pp\to\phi\to\gm\gm$ and 
$pp\to\phi\gm\to 3\gm$ channels at the 13 TeV LHC.

\begin{table}[!ht]
\begin{tabular}{|c|c|c|c|c|c|}
\hline
Category & $2\gm+\geq 0j$ & $2\gm+\geq 1j$ & $2\gm+\geq 2j$ & $\geq 2\gm+\geq 0j$ & $\geq 3\gm+\geq 0j$ \\
\hline
\hline
CMS ($\phi\to\gm\gm$)    & 0.68 & 0.28 & 0.10 & 0.68 & 0.002 \\
\hline
CMS ($\phi\gm\to 3\gm$) & 0.27 & 0.25 & 0.18 & 0.93 & 0.66 \\
\hline
\hline
ATLAS ($\phi\to\gm\gm$)    & 0.48 & 0.19 & 0.07 & 0.48 & 0.001 \\
\hline
ATLAS ($\phi\gm\to 3\gm$) & 0.23 & 0.20 & 0.14 & 0.77 & 0.54 \\
\hline
\end{tabular}
\caption{Cut efficiencies for different selection criteria on the number of selected photons and jets for the 13 TeV ATLAS~\cite{ATLAS:2016eeo} and CMS~\cite{Khachatryan:2016yec} diphoton resonance searches. Here, we use inclusive (up to 2-jets) $pp\to \phi\gm\to 3\gm$ and $pp\to \phi jj\to \gm\gm jj$ processes for the analysis.}
\label{tab:CEhpa}
\end{table}

From Table~\ref{tab:CEhpa}, it is evident that the cut efficiencies are highly dependent on
the selection criteria. For example, if we demand exactly two selected photons and any number of
jets (i.e., $2\gm+\geq 0j$), the cut efficiency is 68\% for CMS ($\phi\to\gm\gm$). On the other hand, this becomes 
28\% when we
select two photons and at least one jet (i.e., $2\gm+\geq 1j$). This is in contrast to the case of 
an $s$-channel production of SM-like
scalar through $gg$ fusion decays to two photons, where we should not expect drastically different cut efficiencies
for $2\gm+\geq 0j$ and $2\gm+\geq 1j$ selection categories. This is because due to the presence of gluons in the
initial state, radiation of jets will be more compared to photon initiated process and therefore in most of the
events one would expect at least one additional jet.

\begin{figure}[!ht]
\subfloat[]{\includegraphics[height=4.5cm,width=4.8cm]{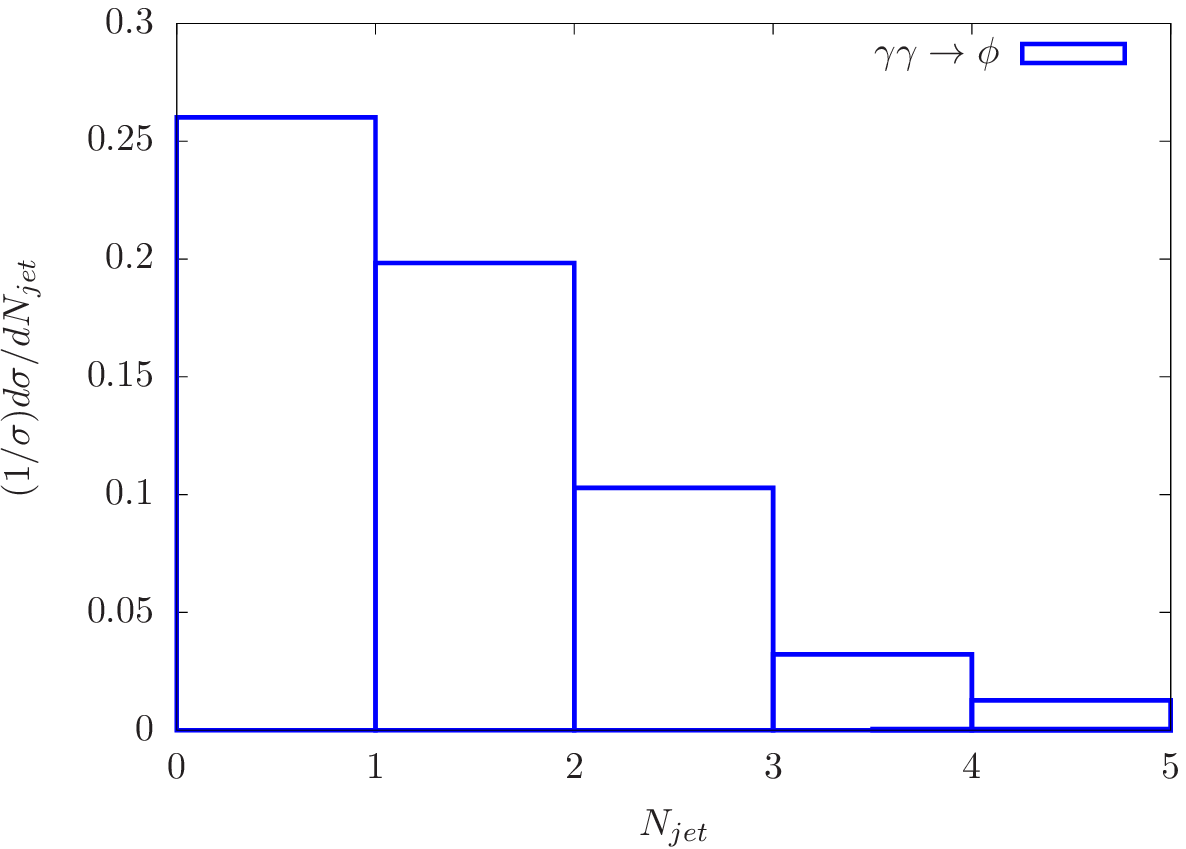}\label{fig:NJ}}
\subfloat[]{\includegraphics[height=4.5cm,width=4.8cm]{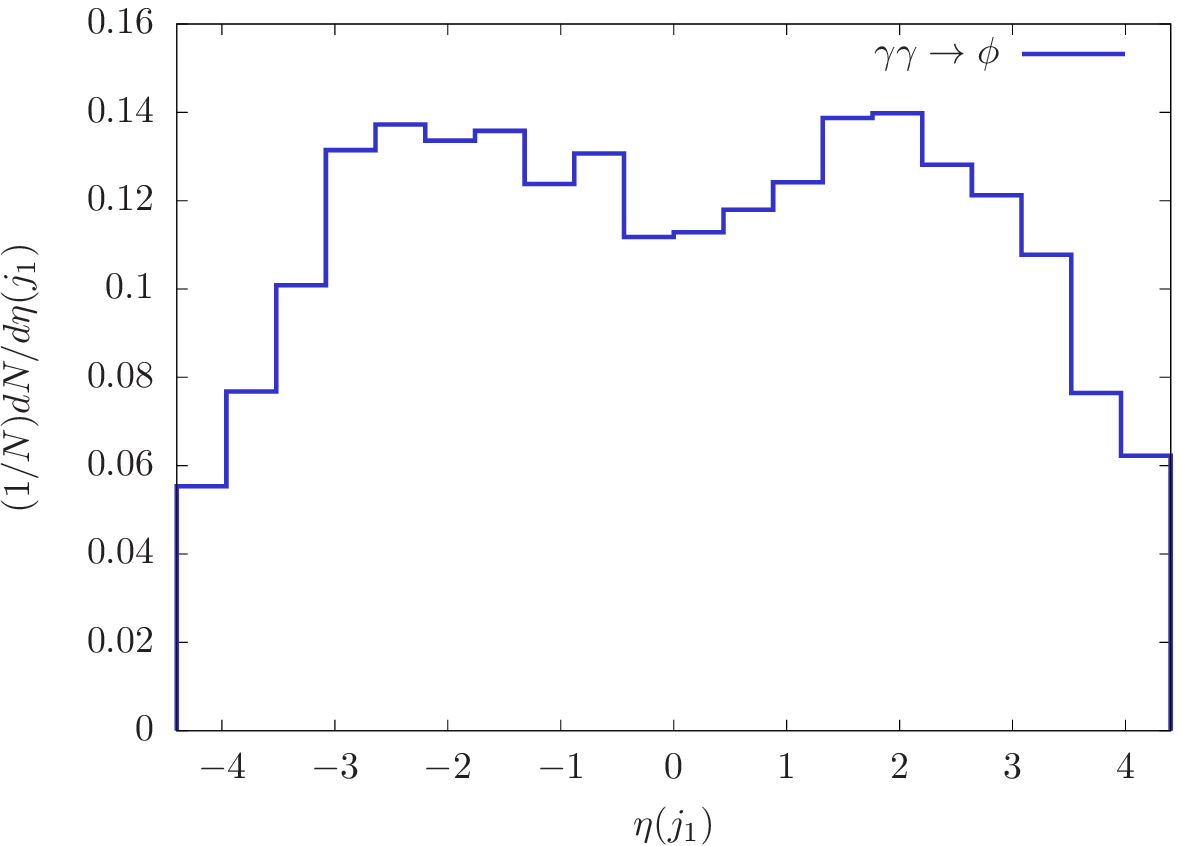}\label{fig:ptj}}
\subfloat[]{\includegraphics[height=4.5cm,width=4.8cm]{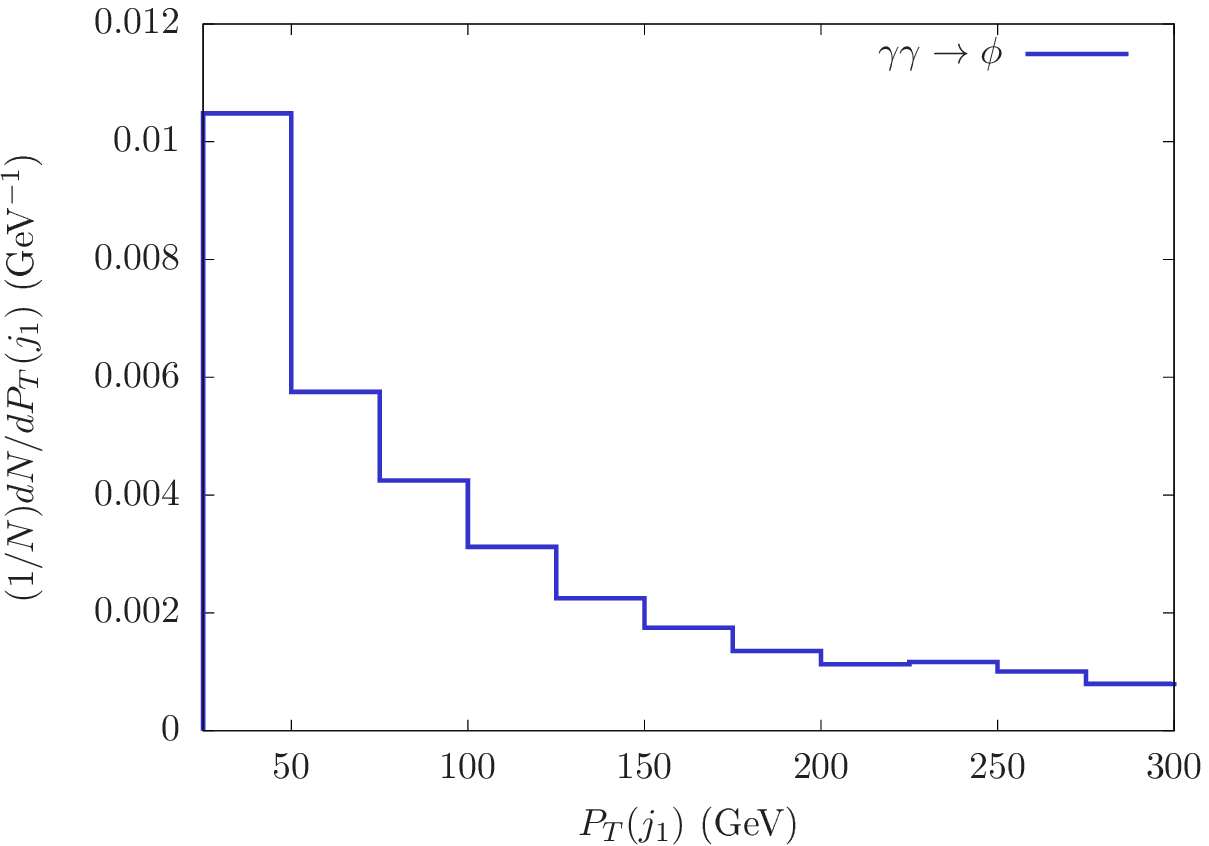}\label{fig:etaj}}
\caption{For the inclusive $pp\to\phi$ process: (a) jet multiplicity distribution for the $2\gm+\geq 0j$ category and, (b) pseudorapidity distribution
and (c) transverse momentum distribution for the hardest jet for $2\gm+\geq 1j$ category. These distributions are drawn after selecting events by applying 13 TeV ATLAS cuts as defined earlier.}
\label{fig:dist}
\end{figure}

In our model, $\phi$ is dominantly produced from $\gm\gm$ fusion, but in many BSM theories, 
a TeV-scale scalar can be dominantly produced from $gg$ fusion, just like the SM Higgs. 
If it is produced from $gg$ fusion, due to the presence of colored particle in the 
initial state, the jet activities is expected to be different from that in $\gm\gm$ fusion
production. Therefore, some jet observables like jet multiplicity, pseudorapidity, or
transverse momentum distributions of jets can act as good discriminating variables
to distinguish different production mechanisms. In Fig.~\ref{fig:NJ}, we show the
jet multiplicity distribution for $\phi$ produced from $\gm\gm$ fusion for the 
$2\gm+\geq 0j$ selection category. In
Figs.~\ref{fig:ptj} and \ref{fig:etaj}, we show the $\eta$ and $p_T$ distributions of
the hardest jet for the $2\gm+\geq 1j$ category. The issue to distinguish different production mechanisms by analyzing various kinematic distributions are discussed in 
Refs.~\cite{Csaki:2016raa,Gao:2015igz,Bernon:2016dow,Ebert:2016idf,Dalchenko:2016dfa,Harland-Lang:2016vzm,Fuks:2016qkf,Mandal:2016bhl} in connection to the 750 GeV diphoton excess. Since the excess is no longer 
there, we do not pursue this issue here.

Now we turn to the 8 TeV LHC data related to the diphoton resonance searches~\cite{Khachatryan:2015qba,CMS:2015cwa,Aad:2015mna}. All these experiments search for either a spin-0 or spin-2
object produced through $gg$ fusion decaying to two photons.
We collect the observed upper limit on the cross sections and the corresponding efficiencies
for resonance mass around 1 TeV for each experiment. We estimate the
cut efficiencies for the process $pp\to\phi\to\gm\gm$ for these experiments by employing
selection cuts in the detector simulator {\sc Delphes3}. From this information,
one can estimate the lower limit on $\Lm$ from the formula given in Eq.~(\ref{eq:Ns}).
The overall result from ATLAS and CMS set an upper limit of the cross 
section roughly 1 fb for 1 TeV resonance mass. Using this we can extract a lower limit on $\Lm$ for our model. 
This depends, however, on how the data is characterized in terms of number of photons and jets, 
since that affects the relevant cut efficiencies. Without reporting all numerical details 
here, we conclude that for the interpretation of the data as $2\gm + \geq 0j$, we obtain a 
lower limit of $\Lm$ in the range $8.5-9$ TeV for $M_{\phi}=1$ TeV. Choosing instead $2\gm + \geq 1j$ would, however, 
give a lower limit $\Lm$ in the range $4.5-5$ TeV.
There is another relevant experiment at the 8 TeV LHC by ATLAS with $\mc{L}=20.3$ fb$^{-1}$ which is important to mention in this context. 
In~\cite{Aad:2015bua}, a triphoton resonance is searched for by ATLAS. This analysis is limited up to a
resonance mass of 500 GeV. In our model, there is a possibility of a three photon final state originating from
$pp\to \phi\gm\to 3\gm$ (although not a triphoton resonance), and therefore, triphoton resonance
searches can also be used to set limits on our model parameters in the future, if the analysis extends the resonance mass range
beyond 1 TeV.

\begin{figure}[!t]
\subfloat[]{\includegraphics[scale=0.55]{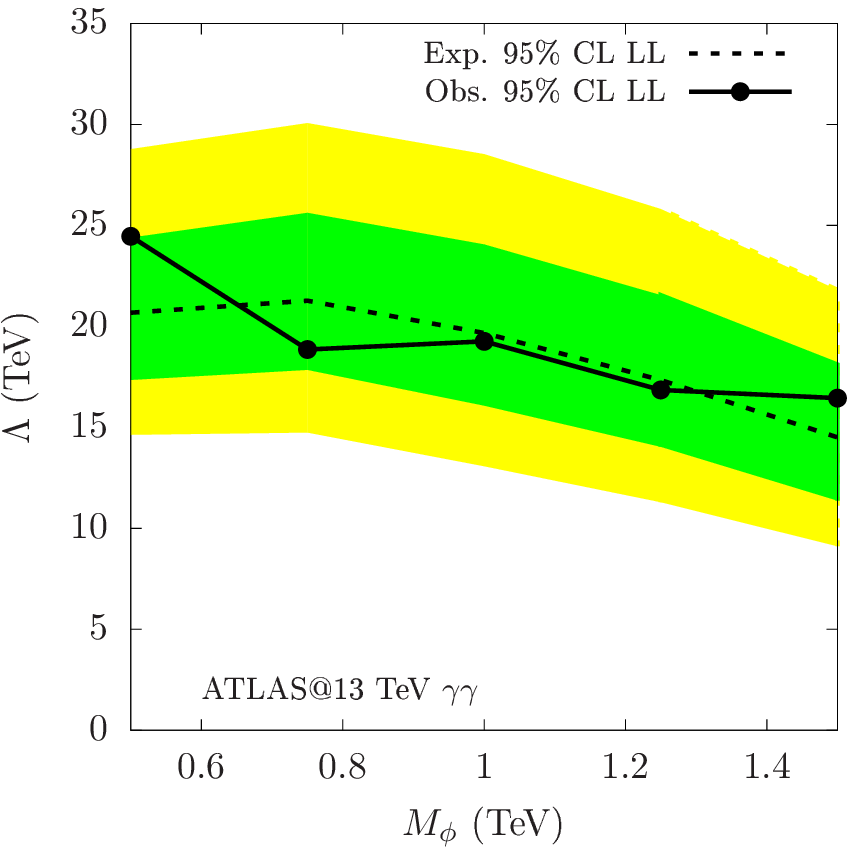}\label{fig:ATLAS}}\hspace{0.25cm}
\subfloat[]{\includegraphics[scale=0.55]{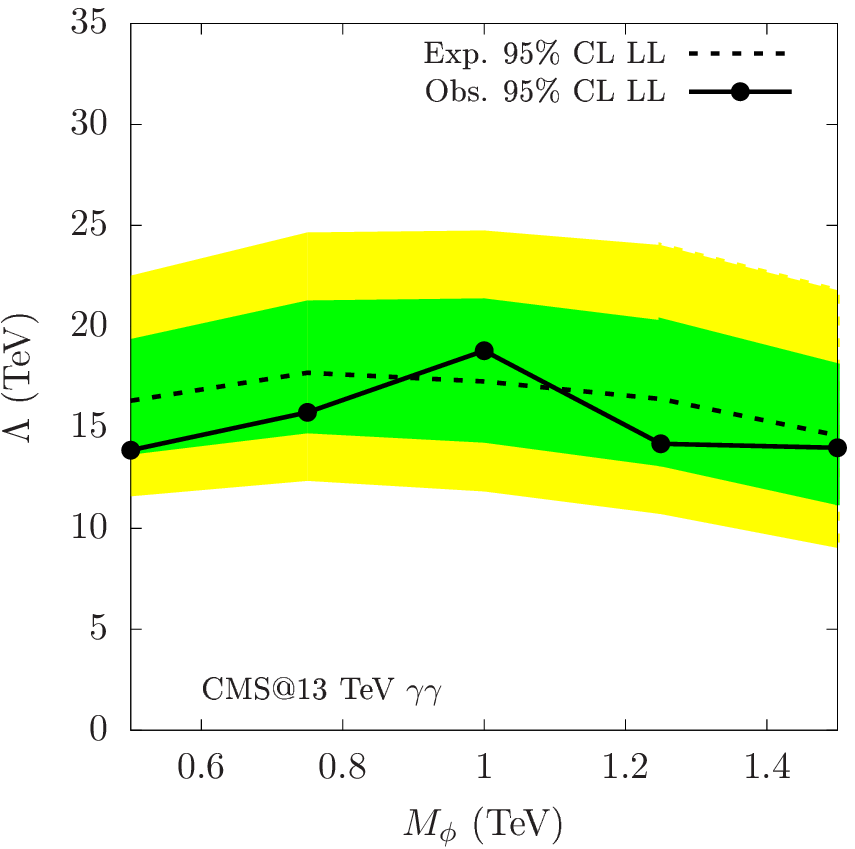}\label{fig:CMS}}\hspace{0.25cm}
\subfloat[]{\includegraphics[scale=0.55]{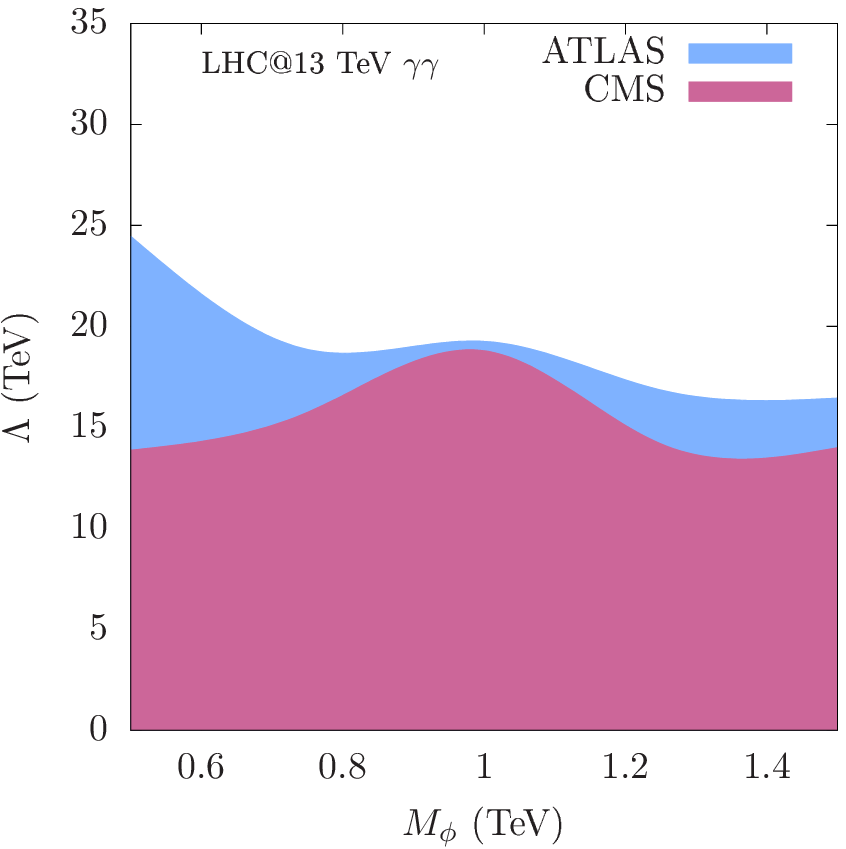}\label{fig:obsexclu}}
\caption{The derived lower limits (LL) on $\Lm$ as functions of $M_{\phi}$ by recasting the upper limit
on $\sg\times \text{BR}$ as set by (a) ATLAS and (b) CMS diphoton resonance searches at the 13 TeV LHC. The 
black dashed and solid curves are expected and observed limits respectively. The green and
yellow bands correspond to $1\sg$ and $2\sg$ uncertainty associated to the expected limits. (c) The shaded
regions are excluded from the observed data.}
\label{fig:exclu}
\end{figure}

The latest 13 TeV LHC data on the diphoton resonance searches also set strong upper limit on 
$\sg\times \text{BR}\sim 1$ fb for a diphoton resonance mass of around 1 TeV. 
Following the same method as used to derive limits
on $\Lm$ from the 8 TeV data, we obtain stronger limits on $\Lm$.
As mentioned earlier, the extraction
of $\Lm$ depends on what selection category is used. For the
category of $2\gm + \geq 0j$ selection for ATLAS and CMS analyses, we get $\Lm\sim 18$ TeV for $M_{\phi}=1$ TeV. 
On the other hand we get slightly smaller $\Lm\sim 12$ TeV for the selection category
$2\gm + \geq 1j$. In Fig.~\ref{fig:exclu}, we show the derived limits on $\Lm$
as functions of $M_{\phi}$ using latest ATLAS and CMS diphoton resonance search data at 13 TeV. 

Finally we wish to briefly discuss the general varying EW theory, which we
introduced in Section~\ref{sec:model}, in light of the LHC data. In 
Table~\ref{tab:PWBR}, we see that the BR for $\phi\to \gm\gm$ is about 65\% for $M_{\phi}=1$ TeV when
only the variation of the EM coupling is considered. For the varying EW theory,
this BR will reduce due to the appearance of other two-body decay modes.
For instance, for the case $\Lm'\gg\Lm$, the BR for $\phi\to\gm\gm$ becomes 39\% (this comes from
$0.65\times 0.6\approx 0.39$ where 0.6 is the $\gm\gm$ BR as shown in Eq.~\eqref{eq:BREW}).
One should note that this BR reduction
does not lower $\Lm$ drastically, since the production cross section scales
as $\Lm^{-2}$. Therefore, the derived $\Lm$, in this case, will reduce by a factor 
$\sqrt{0.6}\approx 0.77$. For example, $\Lm\lesssim 18$ TeV for $M_{\phi}=1$ TeV as shown in 
Fig.~\ref{fig:exclu} will become 14 TeV. 

It is also necessary to check how the latest 13 TeV LHC data on the  
$\gm\gm$, $\gm Z$ and $ZZ$ resonance searches place bounds on $\Lm$ for the varying EW theory. 
The 13 TeV data set rough upper limits on the cross sections for
$\gm\gm$, $\gm Z$
and $ZZ$ resonance searches around mass 1 TeV as 1~fb~\cite{ATLAS:2016eeo,Khachatryan:2016yec}, 10~fb~\cite{CMS:2016pax,CMS:2016cbb} and 20~fb~\cite{ATLAS:2016npe}, respectively.
Since the limits on the cross sections for $\gm Z$ and $ZZ$ resonance searches are
less strong than the $\gm\gm$ limit, they cannot put stronger bounds than that derived from 
$\gm\gm$ data as the $\gm\gm$ BR is the largest for the choice $\Lm'\gg\Lm$ in Eq.~\eqref{eq:BREW}.

\section{Conclusions}
\label{sec:conclusions}

In this paper, we present the first ever study of the phenomenology of a heavy scalar associated with the variation 
of the fine-structure constant. This variation introduces a new scalar field in the theory as originally proposed by Jacob Bekenstein. We introduce a TeV-scale mass of the scalar which
can therefore be accessible at the LHC. This model predicts that the scalar dominantly
couples to photons. Therefore, the dominant production channel is $\gm\gm$ fusion 
and it dominantly decays to a photon pair. 
The scalar can also be produced together with an additional real or virtual photon, which, if virtual, gives rise to a pair of jets or leptons. This gives another prediction: the existence of an additional photon or jets in the events, which are not part of the resonance. These predictions can be searched for in the future at the LHC.

The model we study here is very economical and has only two new parameters, the mass $M_\phi$ of the scalar and the energy scale $\Lambda$. We use latest 13 TeV LHC diphoton resonance search data
to derive exclusion regions on the $M_{\phi}-\Lm$ plane. In particular, for the mass $M_{\phi}\sim 1$
TeV, we obtain the lower limit $\Lm\gtrsim 18$ TeV. We also discuss how different selection
criteria can affect the exclusion limits and derive limits from relevant LHC data for different
selection categories. In this first paper, we primarily consider varying only 
$\al_{\text{EM}}$ in the SM and the resulting phenomenology with photons as clean experimental signals at the LHC. The variations of gauge couplings in the full electroweak theory leads to more complex possibilities, which we will study in a forthcoming paper. Already here, however, we briefly discuss this and derive limits for a specific benchmark point. 

In a broader perspective, one should note that the interaction terms for the new scalar are non-renormalizable such that the theory needs to be UV-completed with new physics at the energy scale $\Lambda$. The scalar in our model could then be interpreted as a moduli field parametrizing a varying electromagnetic coupling. Natural candidates for such new physics are extra dimensions or string theory, with effects that may be within reach of the LHC.

\section*{Acknowledgments}

We thank Elin Bergeaas Kuutmann and Richard Brenner for helpful discussions and the anonymous referees for constructive criticism.
This work is supported by the Swedish Research Council (contracts 621-2011-5107 and 2015-04814) and the Carl Trygger Foundation (contract CTS-14:206).

\bibliography{reference}

\end{document}